# High-modulation-efficiency InGaAsP/Si hybrid MOS optical modulator with Mach–Zehnder interferometer

Jae-Hoon Han, Frederic Boeuf, Shinichi Takagi, and Mitsuru Takenaka

The University of Tokyo, 7-3-1 Hongo, Bunkyo-ku, Tokyo 113-8656, Japan

E-mail: hanjh@mosfet.t.u-tokyo.ac.jp

**A high-modulation-efficiency optical modulator integrated on silicon (Si) is a key enabler for low-power and high-capacity optical interconnects[1–8]. However, Si-based optical modulators suffer from low phase modulation efficiency owing to the weak plasma dispersion effect in Si. Therefore, it is essential to find a novel modulation scheme that is compatible with a Si photonics platform. Here, we demonstrate an InGaAsP/Si hybrid metal-oxide-semiconductor (MOS) optical modulator with a Mach–Zehnder interferometer (MZI) formed by direct wafer bonding with an $Al_2O_3$ bonding interface[9]. Electron accumulation at the InGaAsP MOS interface enables the extraction of the electron-induced refractive index change in InGaAsP, which is significantly greater than that in Si[10,11]. The presented modulator exhibits a phase modulation efficiency of 0.047 Vcm, which is approximately 5 times higher than that of Si MOS optical modulators[12–18]. This approach provides a new efficient scheme of phase modulation on a Si photonics platform for low-power, high-speed, and high-density optical links.**

Si photonics is a promising solution for low-power and large-capacity optical communication[19–21]. In particular, Si optical modulators are essential components for converting electrical signals into optical signals[1]. To achieve optical phase modulation in Si, the free-carrier plasma dispersion effect in Si is a major scheme[22]. Si optical modulators rely on device configurations consisting of pn junctions[23–26], pin junctions[27–29], and metal-oxide-semiconductor (MOS) junctions[12–18] to obtain optical phase modulation through modulation of the carrier density by an electrical signal. Among these configurations, carrier accumulation in an MOS junction consisting of a polycrystalline Si (poly-Si)/SiO$_2$/Si gate stack exhibits the highest phase modulation efficiency with a voltage–length product $V_\pi L$ as small as 0.2 Vcm[14,15]. Thinning a gate oxide thickness is helpful for accumulating more carriers at the same gate voltage, improving the phase modulation efficiency. However, an unavoidable increase in the capacitance of the MOS junction degrades the modulation bandwidth, even though a modulation bandwidth of 40 Gbps has been reported[15]. Hence, it is essential for this inherent trade-off relationship between modulation efficiency and modulation bandwidth to be improved for MOS optical modulators. To overcome the trade-off relationship imposed by the plasma dispersion effect of Si, we focus on using indium gallium arsenide phosphide (InGaAsP) lattice-matched to indium phosphide (InP), which has a large electron-induced refractive index change. To introduce InGaAsP into a Si photonics platform, we have proposed a Si hybrid platform for an MOS optical modulator using direct wafer bonding with high-k dielectrics[9,30]. Recently, a microring tunable laser with an InP/Si hybrid MOS structure using direct wafer bonding has been reported[31], for which direct wavelength tuning was achieved by accumulated carriers at the InP/Al$_2$O$_3$/Si MOS structure[31], suggesting the usefulness of a Si hybrid MOS structure in optical modulation. However, the phase modulation efficiency has not yet been clarified for the case that a Si hybrid MOS structure is used with a Mach–Zehnder interferometer (MZI) configuration to build an optical modulator.

In this study, we investigated an MZI optical modulator with an InGaAsP/Si hybrid

MOS phase shifter as shown in Fig. 1a. Thin III-V layers containing an InGaAsP layer were bonded on a Si waveguide to form an InGaAsP/Al$_2$O$_3$/Si MOS capacitor, which operates as a phase shifter in an MZI interferometer. To achieve a smooth transition from a Si waveguide to the phase shifter, a tapered structure was exploited (Supplementary Section I). The cross-sectional structure of the MOS phase shifter is shown in Fig. 1b. The total thickness of the Si rib waveguide on the SiO$_2$ buried oxide (BOX) was 220 nm, which is the standard thickness of a Si photonics platform operating at a 1.55 µm wavelength. To form the taper, the width of the Si waveguide was designed to be 1 µm. A 110-nm-thick $n$-type InGaAsP layer with a 50-nm-thick InP layer was bonded on the Si waveguide with an Al$_2$O$_3$ bonding interface, which also acts as a gate oxide. The mode profile of the fundamental transverse-electric (TE) mode is overlaid in Fig. 1b. Note that the thin III-V layers contribute to the enhancement in the optical confinement around the bonding interface, resulting in further improvement in the phase modulation efficiency (Supplementary Section II). When a gate voltage is applied between the $n$-type InGaAsP and $p$-type Si layers, electrons accumulate at the InGaAsP interface, which contributes to optical phase modulation. The equivalent oxide thickness (EOT) of the Al$_2$O$_3$ gate oxide, which is the equivalent SiO$_2$ thickness producing the same gate capacitance, was assumed to be 5 nm in the following numerical analyses. It is worth noting that the presented hybrid MOS structure also enables low series resistance and low optical loss by replacing poly-Si with high-electron-mobility crystalline InGaAsP. The low series resistance reduces the resistive-capacitance delay, improving the modulation bandwidth (Supplementary Section III). Hence, an InGaAsP/Si hybrid MOS optical modulator can simultaneously achieve high a modulation efficiency, low insertion loss, and high modulation bandwidth.

To estimate the modulation efficiency of an InGaAsP/Si hybrid MOS optical modulator,

the carrier-induced changes in the refractive index and absorption of InGaAsP[10,11] were calculated (for details see Supplementary Section IV). Since the plasma dispersion effect is inversely proportional to the effective mass according to the Drude model, the light electron effective mass in InGaAsP results in a greater change in the refractive index change than that in Si. Additionally, in contrast to Si, the band-filling effect and bandgap shrinkage contribute to the free-carrier-induced changes in the refractive index and absorption of $n$-type InGaAsP[10,11,32]. By changing the stoichiometry of InGaAsP, we can tune its optical and electrical properties while keeping its lattice constant matched to that of InP[33–35]. Generally speaking, InGaAsP with a high As composition has large carrier-induced effects, while its bandgap is narrowed; thus, we have to determine the As composition to maximize the carrier-induced effects without band-edge absorption at a 1.55 µm wavelength. Thus, we chose $In_{0.68}Ga_{0.32}As_{0.7}P_{0.3}$, whose bandgap energy was 0.91 eV ($\lambda_g = 1.37$ µm). By taking into account the plasma dispersion effect, band-filling effect, and bandgap shrinkage, we calculated the electron-induced changes in the refractive index and absorption of InGaAsP as shown in Figs. 2a and 2b, respectively (for details see Supplementary Fig. S7). In addition to Si as a reference, we also calculated the free-carrier effects in InP. As shown in Fig. 2a, InGaAsP exhibited a greater electron-induced refractive index change than Si and InP. Owing to the band-filling effect, the electron-induced refractive index change was not linear to the electron density (See Supplementary Fig. S7a). When the electron density was around $2 \times 10^{18}$ cm$^{-3}$, the refractive index change of InGaAsP was 17 times greater than that of Si. Since InGaAsP has a lighter electron effective mass than InP, the refractive index change in InGaAsP is expected to be greater than that in InP. Meanwhile, the electron-induced absorption change of InGaAsP was approximately half of that of Si because of the light electron effective mass and high electron mobility in InGaAsP, as shown in Fig. 2b. Although hole-induced absorption in InGaAsP due to the inter-valence band absorption was significantly greater than that in Si (Supplementary Fig. S8b), the $n$-InGaAsP/$p$-Si hybrid MOS phase shifter did not induce holes in the InGaAsP layer, enabling a large phase shift with a small absorption change. Thus, the InGaAsP/Si hybrid MOS phase shifter is suitable for not only optical modulators but also low-crosstalk optical switches[36] (See also Supplementary Fig. S9).

Using the calculated free-carrier effects in InGaAsP (Figs. 2a and 2b), the phase modulation efficiency of an InGaAsP/Si hybrid MOS optical modulator was numerically analyzed (for details see Supplementary Section V). The impurity concentration of each semiconductor was assumed to be $10^{18}$ cm$^{-3}$. Figure 2c shows the changes in the effective refractive index of InGaAsP/Si, InP/Si hybrid, and Si MOS phase shifters. As shown in Fig. 2c, the InGaAsP device exhibited an approximately two times and eight times larger refractive index change than the InP and Si devices, respectively. We also calculated the voltage–length product $V_\pi L$ of the InGaAsP/Si hybrid MOS optical modulator and its carrier-induced loss as a function of the gate voltage $V_g$, as shown in Fig. 2d. The expected $V_\pi L$ at $V_g$ of 1 V was 0.051 Vcm, which was approximately 6 times lower than that of a reported Si MOS optical modulator with the same EOT of 5 nm[17]. From the results in Fig. 2d, the insertion loss at 0 V was 2.3 dB/mm and the attenuation increase at 1 V was 0.45 dB/mm. From these results, $\alpha V_\pi L$, which is the product of the propagation loss per unit length $\alpha$ and $V_\pi L$, was estimated to be 1.2 dBV, which was significantly smaller than the reported values of 7 dBV for a Si MOS optical modulator[18], 9.5 dBV for a pin-junction optical modulator[29], and 25.4 dBV for a pn-junction optical modulator[23]. Since there is a trade-off relationship between the modulation efficiency and optical loss in Si-based optical modulators, the improvement in $\alpha V_\pi L$ by the InGaAsP/Si hybrid MOS phase shifter is essential for high-efficiency and low-loss optical modulators.

To demonstrate the expected high-efficiency phase modulation with low insertion loss, we fabricated an InGaAsP/Si hybrid MOS optical modulator with an MZI using void-free wafer bonding with an $Al_2O_3$ bonding interface (see Supplementary Section VI)[9]. Figure 3a is a plan-view scanning electron microscope image of the hybrid MOS phase shifter and Si waveguide with a 3 dB coupler. The taper structure can be clearly observed in the inset of Fig. 3a. A cross-sectional transmission electron microscope image of the wafer-bonded InGaAsP/Si hybrid MOS phase shifter is shown in Fig. 3b. The III-V layer was uniformly bonded on Si with the $Al_2O_3$ gate oxide as shown in the inset of Fig. 3b. First, we evaluated the electrical properties of the wafer-bonded InGaAsP/$Al_2O_3$/Si hybrid MOS capacitor, which are important for achieving efficient phase modulation. The capacitance–voltage ($C$–$V$) characteristic of the hybrid MOS capacitor in Fig. 3c showed a change in capacitance due to carrier depletion with no frequency dispersion from 10 kHz to 1 MHz, indicating superior MOS interfacial quality. From the accumulated capacitance, the EOT of the wafer-bonded capacitor was extracted to be 5.6 nm. Then, we evaluated the phase modulation of the InGaAsP/Si hybrid MOS phase shifter by employing an asymmetric MZI with a 20 µm difference in the optical path length between the two MZI arms. Figure 3d shows the transmission spectra of the InGaAsP/Si hybrid MOS optical modulator with an asymmetric MZI. The phase shifter length ($L$) was 500 µm. When the gate voltage was increased from 0 to 0.8 V, a shift of the resonance wavelength peak of 13.6 nm, which was associated with a phase shift, was observed as shown in Fig. 3d. From the shift of the peak wavelength in Fig. 3d, the phase shift was extracted as shown in Fig. 3e. The phase shift increased linearly with gate voltage because of the carrier accumulation of the MOS capacitor. From the slope in Fig. 3e, the gate voltage for a $\pi$ phase shift ($V_\pi$) was found to be 0.86 V, resulting in a voltage–length product ($V_\pi L$) of 0.047 Vcm, which was in good agreement with the simulation result. Figure 3f shows the increase in attenuation with the gate voltage obtained by measurement (see Supplementary Section VII). The numerical result is shown as a dashed line in Fig. 3f, which was in good agreement with the measurement. An extremely low attenuation increase of 0.23 dB was estimated when $\pi$ phase shift was induced, while the attenuation increase at $\pi$ phase shift of Si MOS optical modulator is over 2.3 dB[15]. Thus, we successfully demonstrated an extremely high-efficient phase modulation with a low change in attenuation using an InGaAsP/Si hybrid MOS phase shifter.

Figure 4a shows the benchmark voltage–length product ($V_\pi L$) of MOS optical modulators as a function of EOT. Thanks to the large electron-induced refractive index change in InGaAsP, the InGaAsP/Si hybrid MOS optical modulator exhibited at least 5 times lower $V_\pi L$ than that of a Si MOS optical modulator with the same EOT[16]. This means that the phase modulation efficiency can be improved without an increase in the gate capacitance by EOT scaling. Thus, the trade-off relationship between the phase modulation efficiency and the modulation bandwidth of MOS optical modulators can be essentially improved by introducing the InGaAsP/Si hybrid MOS phase shifter into a Si photonics platform. The modulation bandwidth, which is limited by the resistive-capacitance delay, is expected to be further improved by reducing the series resistance. Owing to the high electron mobility in InGaAsP, $n$-type InGaAsP has more than 32 times lower resistivity than $n$-type Si with the same doping concentration (Supplementary Section III), enabling a modulation bandwidth of more than 50 Gbps with better modulation efficiency than that of state-of-the-art Si MOS optical modulators[14,15].

The optical insertion loss can also be reduced by replacing $n$-type poly-Si with $n$-type crystalline InGaAsP. Si MOS optical modulators[12–18] typically use poly-Si with a doping concentration as high as $10^{18}$ cm$^{-3}$ to reduce the series resistance for achieving a modulation bandwidth over 25 Gbps[14,15,17,18], causing a high propagation loss due to free-carrier absorption. Furthermore, the scattering at grain boundaries in poly-Si results in a propagation loss of more than 1.0 dB/mm[17]. Figure 4b shows the relationship between $V_\pi L$ and propagation loss. The simulated results are shown as dashed lines, which are in good agreement with the experimental results (see also Supplementary Section VII). Si MOS optical modulators typically exhibit a product of propagation loss and phase modulation efficiency ($\alpha V_\pi L$) of over 7 dBV[18]. Since the electron-induced absorption in InGaAsP is half of that in Si and because of its high modulation efficiency, the InGaAsP/Si hybrid MOS optical modulator exhibits $\alpha V_\pi L$ of 1.4 dBV even with the same doping concentration of $10^{18}$ cm$^{-3}$, as shown in Fig. 4b. Owing to the 10 times lower resistivity of InGaAsP with $10^{17}$ cm$^{-3}$ doping concentration than that of Si with a doping concentration of $10^{18}$ cm$^{-3}$ (Supplementary Section III), a further reduction of $\alpha V_\pi L$ without bandwidth degradation is expected by reducing the doping concentration of InGaAsP. The attenuation increase at $\pi$ phase shift is also an important benchmark of optical modulators. Figure 4c shows the attenuation increase at $\pi$ phase shift of InGaAsP/Si hybrid and Si MOS phase shifters as a function of the phase shifter length for $\pi$ phase shift ($L_\pi$). The solid line shows the numerically calculated results. Since the hole-induced refractive index change in Si is not proportional to the hole density[22], the attenuation change was dependent on $L_\pi$. When $L_\pi$ was 1000 μm, the attenuation increase in the Si MOS optical modulator was approximately 2.3 dB, which degrades the optical modulation amplitude (OMA) as we will discuss later. In contrast, InGaAsP exhibited an extremely small attenuation increase of 0.23 dB at the 500-μm-long phase shifter, which was approximately 10 times smaller than that of the Si MOS optical modulator (for details see Supplementary Section VII). Furthermore, the small dependence of the attenuation increase of the InGaAsP/Si hybrid MOS optical modulator on $L_\pi$ enables the miniaturization of the device, while the attenuation increase of the Si MOS optical modulator is significantly dependent on $L_\pi$, as shown in Fig. 4c. Thus, an InGaAsP/Si hybrid MOS optical modulator allows the phase modulation efficiency and insertion loss to be simultaneously improved, which is essential for optical modulators.

The dynamic modulation characteristics including eye diagram of an InGaAsP/Si hybrid MOS optical modulator were also numerically analyzed to clarify the impact of the introduction of the InGaAsP/Si hybrid MOS structure on the modulation performances such as the modulation bandwidth, extinction ratio (ER), and OMA. To make a direct comparison to state-of-the-art Si MOS optical modulators, we assumed the device structure reported in Refs. 14 and 15 (See Figure S13). We developed the semi-analytical model to calculate the modulation characteristics, which enabled us to reproduce the modulation efficiency and bandwidth of the Si MOS optical modulator in Refs. 14 and 15 (for details see Supplementary Section VIII). Figure 5a shows the relationships between the modulation efficiency and cut-off frequency of an InGaAsP/Si hybrid MOS optical modulator and a Si MOS optical modulator where their access resistance was assumed to be 4 Ωmm. The cut-off frequency was calculated from the resistive-capacitance delay. As shown in the figure, the InGaAsP/Si hybrid MOS modulator exhibits simultaneously better modulation efficiency and cut-off frequency than those of the Si MOS modulator, meaning the improvement in the trade-off relationship between the modulation efficiency and bandwidth as discussed in the previous paragraph. The OMA, which is defined as the optical power difference between an on and off states, is another figure-of-merit of optical modulators, reflecting the opening of eye diagram in terms of ER and insertion loss. We assumed a voltage swing of 1 V and 0 dBm input power for OMA calculation. To consider the modulation bandwidth simultaneously, we defined dynamic OMA by

$$OMA/\sqrt{1 + f_0^2/f_c^2}, \tag{1}$$

where $f_c$ is the cut-off frequency of the modulation and $f_0$ is the minimum operation frequency of the system, i.e. half of baud rate. Figure 5b show a comparison of the dynamic OMA between InGaAsP/Si MOS and Si MOS optical modulators as a function of EOT. The EOT should be increased to improve the modulation bandwidth, while a decrease in the modulation efficiency causes an increase in the length of the phase shifter. Thus, in case of the Si device, the dynamic OMA cannot be improved even when the EOT increases because an increase in the insertion loss degrades the OMA significantly. As shown in Fig. 5b, the dynamic OMA of the Si MOS optical modulator was worse than -10 dBm regardless of EOT. On the other hand, the InGaAsP/Si hybrid MOS optical modulator exhibited much better dynamic OMA than that of the Si device. Moreover, the deterioration of the dynamic OMA with an increase in EOT was very gentle owing to the small absorption. Thus, we can achieve the dynamic OMA of approximately -2.5 dBm when the EOT is greater than 10 nm, meaning that the InGaAsP/Si hybrid MOS optical modulator is suitable for high-speed modulation. To show the impact of the dynamic OMA on the dynamic modulation characteristics, we simulated eye diagrams of the Si and III-V/Si hybrid MOS optical modulators. For the analysis, we assumed the EOT of 14 nm which maximizes the dynamic OMA of the III-V/Si hybrid device. Figures 5c and 5d shows eye diagrams of the Si and III-V/Si hybrid devices modulated with a 53 Gb/s non-return-to-zero (NRZ) signal, respectively. The Si modulator exhibited a poor eye opening because of the worse OMA of -15 dBm and ER of 1.6 dB. On the other hand, the III-V/Si hybrid modulator exhibited a clear eye diagram with the OMA of -4.5 dBm and ER of 12 dB, which were significantly better than those of the Si device. The impact of the dynamic OMA becomes more prominent in a multi-level modulation format. Figures 5e and 5f are eye diagrams of the Si and III-V/Si hybrid optical modulators modulated with a 53-Gbaud/s 4-level pulse-amplitude modulation (PAM-4), respectively. In the case of the Si device, the eye pattern collapsed because of its poor OMA of approximately -19 dBm. To improve the eye opening, we should increase the input optical power, resulting in an increase in the power consumption significantly. On the other hand, we can achieve a clear eye opening with the OMA of approximately -9.5 dBm and outer ER of 13.8 dB by using the III-V/Si hybrid MOS structure. Thus, the III-V/Si hybrid MOS optical modulator is much more advantageous than the Si MOS optical modulator for high-speed multi-level modulation which is going to be a standard of optical interconnection.

In conclusion, we successfully demonstrated a high-efficiency MZI optical modulator with an InGaAsP/Si hybrid MOS phase shifter using direct wafer bonding. We achieved a phase modulation efficiency of 0.047 Vcm, which is, to the best of our knowledge, one of the best values reported for semiconductor-based optical modulators. The presented modulator needs fewer charges for $\pi$ phase shift than Si-based optical modulators, enabling low optical insertion loss simultaneously. Owing to these excellent performance, its dynamic OMA is also significantly improved compare to conventional Si MOS. Thus, the InGaAsP/Si hybrid MOS phase shifter is expected to be particularly useful for optical modulators with advanced modulation formats and optical switches with high modulation bandwidth. Since various semiconductors including SiGe, Ge, and other III-V semiconductors can be integrated using the presented scheme, the hybrid MOS structures are expected to provide many applications for near- and mid-infrared wavelengths.

**References**

1. Reed, G. T., Mashanovich, G., Gardes, F. Y. & Thomson, D. J. Silicon optical modulators. *Nat. Photonics* **4**, 518–526 (2010).

2. Kuo, Y.-H., Lee, Y. K., Ge, Y., Ren, S., Roth, J. E., Kamins, T. I., Miller, D. A. B. & Harris, J. S. Strong quantum-confined Stark effect in germanium quantum-well structures on silicon. *Nature* **437**, 1334 (2005).

3. Jacobsen, R. S., Andersen, K. N., Borel, P. I., Fage-Pedersen, J., Frandsen, L. H., Hansen, O., Kristensen, M., Lavrinenko, A. V., Moulin, G., Ou, H., Peucheret, C., Zsigri, B. & Bjarklev, A. Strained silicon as a new electro-optic material. *Nature* **441**, 199 (2006).

4. Liu, J., Beals, M., Pomerene, A., Bernardis, S., Sun, R., Cheng, J., Kimerling, L. C. & Michel, J. Waveguide-integrated, ultralow-energy GeSi electro-absorption modulators. *Nat. Photonics* **2**, 433 (2008).

5. Liu, M., Yin, X., Ulin-Avila, E., Geng, B., Zentgraf, T., Ju, L., Wang, F. & Zhang, X. A graphene-based broadband optical modulator. *Nature* **474**, 64 (2011).

6. Kim, Y., Takenaka, M., Osada, T., Hata, M. & Takagi, S. Strain-induced enhancement of plasma dispersion effect and free-carrier absorption in SiGe optical modulators. *Sci. Rep.* **4**, 4683 (2014).

7. Phare, C. T., Lee, Y.-H. D., Cardenas, J. & Lipson, M. Graphene electro-optic modulator with 30 GHz bandwidth. *Nat. Photon.* **9**, 511 (2015).

8. Sun, C. *et al.* Single-chip microprocessor that communicates directly using light.

*Nature* **528**, 534 (2015).

9. Han, J.-H., Takenaka, M., Takagi, S. Study on void reduction in direct wafer bonding using $Al_2O_3$/$HfO_2$ bonding interface for high-performance Si high-k MOS optical modulators. *Jpn. J. Appl. Phys.* **55**, 04EC06 (2016).

10. Bennett, B. R., Soref, R. A. & Del Alamo, J. A. Carrier-induced change in refractive index of InP, GaAs, and InGaAsP. *IEEE J. Quantum Electron.* **26**, 113–122 (1990).

11. Weber, J.-P. Optimization of the carrier-induced effective index change in InGaAsP waveguides-application to tunable Bragg filters. *IEEE J. Quantum Electron.* **30**, 1801–1816 (1994).

12. Liu, A., Jones, R., Liao, L., Samara-Rubio, D., Rubin, D., Cohen, O., Nicolaescu, R. & Paniccia, M. A high-speed silicon optical modulator based on a metal–oxide–semiconductor capacitor. *Nature* **427**, 615 (2004).

13. Liao, L., Samara-Rubio, D., Morse, M., Liu, A., Hodge, D., Rubin, D., Keil, U. D. & Franck, T. High speed silicon Mach-Zehnder modulator. *Opt. Express* **13**, 3129 (2005).

14. Webster, M. *et al.* An efficient MOS-capacitor based silicon modulator and CMOS drivers for optical transmitters. *In Int. Conf. Group IV Photonics* Paper WB1 (IEEE, 2014).

15. Webster, M., *et al.* Silicon Photonic Modulator Based on a MOS-Capacitor and a CMOS Driver. *In Compound Semiconductor Integrated Circuit Symp.* Paper E.3 (IEEE, 2014).

16. Campenhout, J. V., Pantouvaki, M., Verheyen, P., Selvaraja, S., Lepage, G., Yu, H., Lee, W., Wouters, J., Goossens, D., Moelants, M., Bogaerts, W. & Absil, P. Low-Voltage, Low-Loss, Multi-Gb/s Silicon Micro-Ring Modulator based on a MOS Capacitor. *In Optical Fiber Communication Conf.* Paper OM2E.4 (2012).

17. Fujikata, J., Takahashi, M., Takahashi, S., Horikawa, T. & Nakamura, T. High-speed and high-efficiency Si optical modulator with MOS junction, using solid-phase crystallization of polycrystalline silicon. *Jpn. J. Appl. Phys.* **55**, 042202 (2016).

18. Fujikata, J., Takahashi, S., Takahashi, M., Noguchi, M., Nakamura, T. & Arakawa, Y. High-performance MOS-capacitor-type Si optical modulator and surface-illumination-type Ge photodetector for optical interconnection. *Jpn. J. Appl. Phys.* **55**, 04EC01 (2016).

19. Soref, R. The Past, Present, and Future of Silicon Photonics. *IEEE J. Sel. Top. Quantum Electron.* **12**, 1678–1686 (2006).

20. Beausoleil, R. G., McLaren, M. & Jouppi, N. P. Photonic architectures for high-performance data centers. *IEEE J. Sel. Top. Quantum Electron.* **19**, 3700109 (2013).

21. Taubenblatt, M. A. Optical interconnects for high-performance computing. *J. Lightwave Technol.* **30**, 448–457 (2012).

22. Soref, R. & Bennett, B. Electrooptical effects in silicon. *IEEE J. Quantum Electron.* **23**, 123–129 (1987).

23. Tu, X., Liow, T.-Y., Song, J., Yu, M. & Lo, G. Q. Fabrication of low loss and high speed silicon optical modulator using doping compensation method. *Opt. Express* **19**, 18029 (2011).

24. Baehr-Jones, T., Ding, R., Liu, Y., Ayazi, A., Pinguet, T., Harris, N. C., Streshinsky, M., Lee, P., Zhang, Y., Lim, A. E.-J., Liow, T.-Y., Teo, S. H.-G., Lo, G.-Q. & Hochberg, M. Ultralow drive voltage silicon traveling-wave modulator. *Opt. Express* **20**, 12014 (2012).

25. Reed, G. T., Mashanovich, G. Z., Gardes, F. Y., Nedeljkovic, M., Hu, Y., Thomson, D. J., Li, K., Wilson, P. R., Chen, S.-W. & Hsu, S. S. Recent breakthroughs in carrier depletion based silicon optical modulators. *Nanophotonics* **3**, 229 (2014).

26. Reed, G. T., Thomson, D. J., Gardes, F. Y., Hu, Y., Fedeli, J.-M. & Mashanovich, G. Z. High-speed carrier-depletion silicon Mach-Zehnder optical modulators with lateral PN junctions. *Front. Phys.* **2**, 77 (2014).

27. Xu, Q., Schmidt, B., Pradhan, S. & Lipson, M. Micrometer-scale silicon electro-optic modulator. *Nature* **435**, 325 (2007).

28. Green, W. M. J., Rooks, M. J., Sekaric, L. & Vlasov, Y. A. Ultra-compact, low RF power, 10 Gb/s silicon Mach-Zehnder modulator. *Opt. Express* **15**, 17106 (2012).

29. Akiyama, S. & Usuki, T. High-speed and efficient silicon modulator based on forward-biased pin diodes. *Front. Phys.* **2**, 65 (2014).

30. Han, J.-H, Takenaka, M. & Takagi, S. Feasibility study of III-V/Si hybrid MOS optical modulators consisting of $n$-InGaAsP/$Al_2O_3$/$p$-Si MOS capacitor formed by wafer bonding. *In Int. Conf. Group IV Photonics* Paper ThP16 (IEEE, 2016).

31. Liang, D., Huang, X., Kurczveil, G., Fiorentino, M. & Beausoleil, R. G. Integrated finely tunable microring laser on silicon. *Nat. Photon.* 163 (2016).

32. Botteldooren, D. & Baets, R. Influence of band-gap shrinkage on the carrier-induced refractive index change in InGaAsP. *Appl. Phys. Lett.* **54**, 1989 (1989).

33. Adachi, S. Material parameters of $In_{1-x}Ga_xAs_yP_{1-y}$ and related binaries. *J. Appl. Phys.* **53**, 8775 (1982).

34. Adachi, S. Optical dispersion relations for GaP, GaAs, GaSb, InP, InAs, InSb, $Al_xGa_{1-x}As$, and $In_{1-x}Ga_xAs_yP_{1-y}$. *J. Appl. Phys.* **66**, 6030 (1989).

35. Sotoodeh, M., Khalid, A. H. & Rezazadeh, A. A. Empirical low-field mobility model for III–V compounds applicable in device simulation codes. *J. Appl. Phys.* **87**, 2890 (2000).

36. Ikku, Y., Yokoyama, M., Noguchi, N., Ichikawa, O., Osada, T., Hata, M., Takenaka, M. & Takagi, S. Low-crosstalk $2 \times 2$ InGaAsP photonic-wire optical switches using III-V CMOS photonics platform. *In Eur. Conf. on Optical Communication* Paper P.2.19 (2013).

**Acknowledgements**

This work was supported in part by the NEDO PECST project and a Grant-in-Aid for Young Scientists (A) from MEXT. The authors would like to thank Drs. O. Ichikawa, M. Yokoyama, T. Yamamoto, and H. Yamada in Sumitomo Chemical Corporation for their collaboration. J.-H. Han would also like to thank JSPS for a research fellowship.

**Author contributions**

J.-H. H. contributed to the idea, simulation, fabrication, measurement, and manuscript preparation. F. B. contributed to the simulation, manuscript preparation and discussion. S. T. contributed to the discussion and high-level project supervision. M. T. contributed to the idea, discussion, manuscript revision, and high-level project supervision.

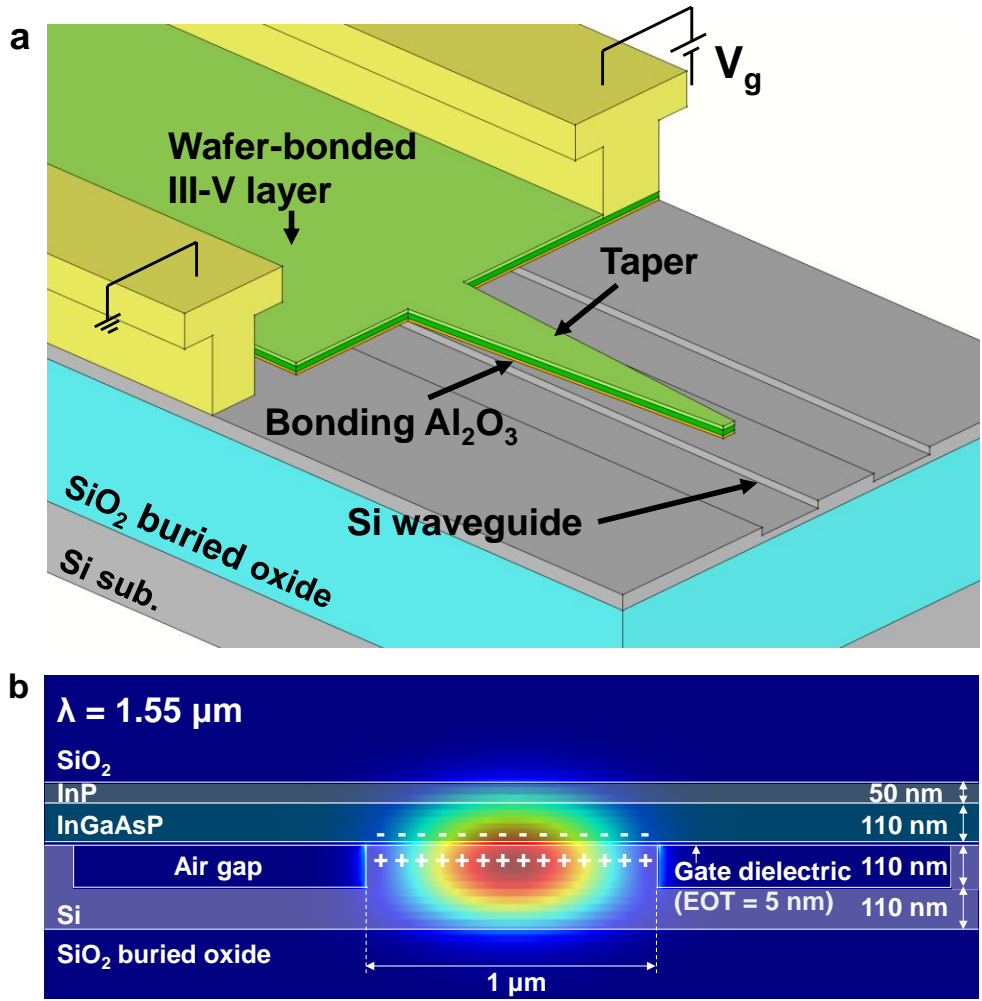

Fig. 1. **Schematics of an InGaAsP/Si hybrid MOS optical modulator. a,** 3D view of the phase shifter. **b,** Cross-sectional view of the phase shifter and its fundamental TE mode. The InGaAsP/InP layers are bonded on a Si waveguide.

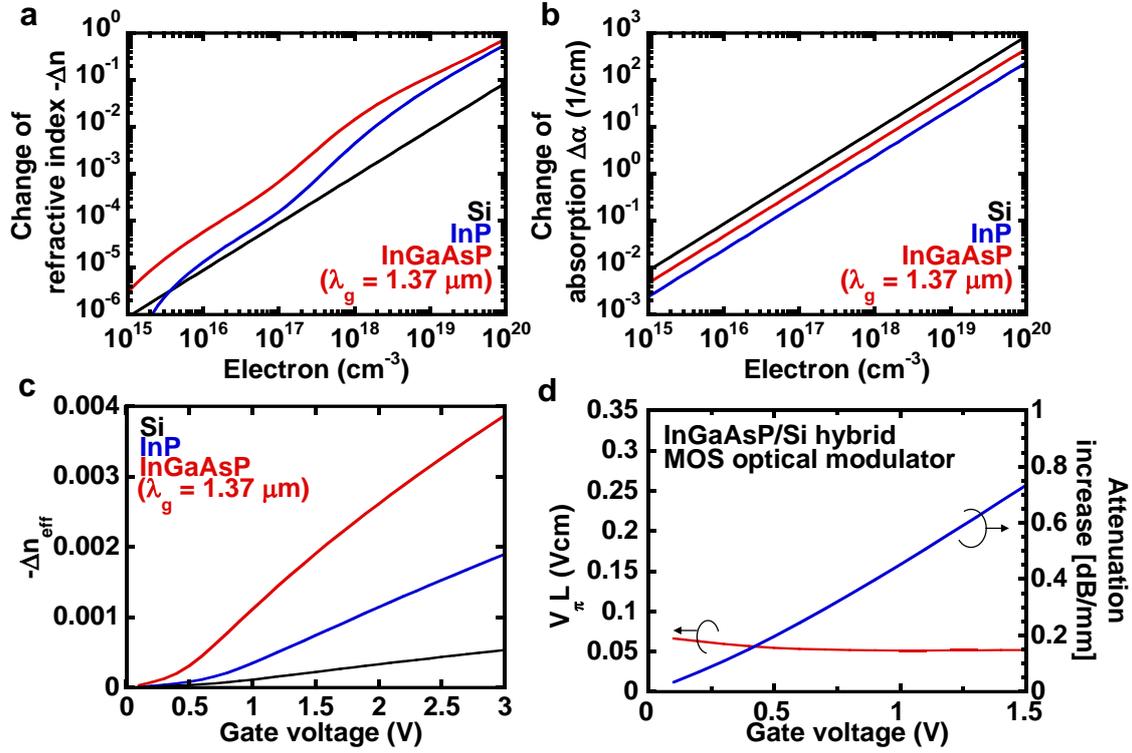

Fig. 2. **Numerical analysis of an InGaAsP/Si hybrid MOS optical modulator.** (a) Electron-induced refractive index changes of InGaAsP, InP, and Si. (b) Electron-induced absorption changes of InGaAsP, InP, and Si. (c) Calculated effective refractive index change of InGaAsP/Si hybrid, InP/Si hybrid, and Si MOS optical modulators. (d) Simulated $V_\pi L$ and attenuation increase of an InGaAsP/Si hybrid MOS optical modulator as a function of gate voltage.

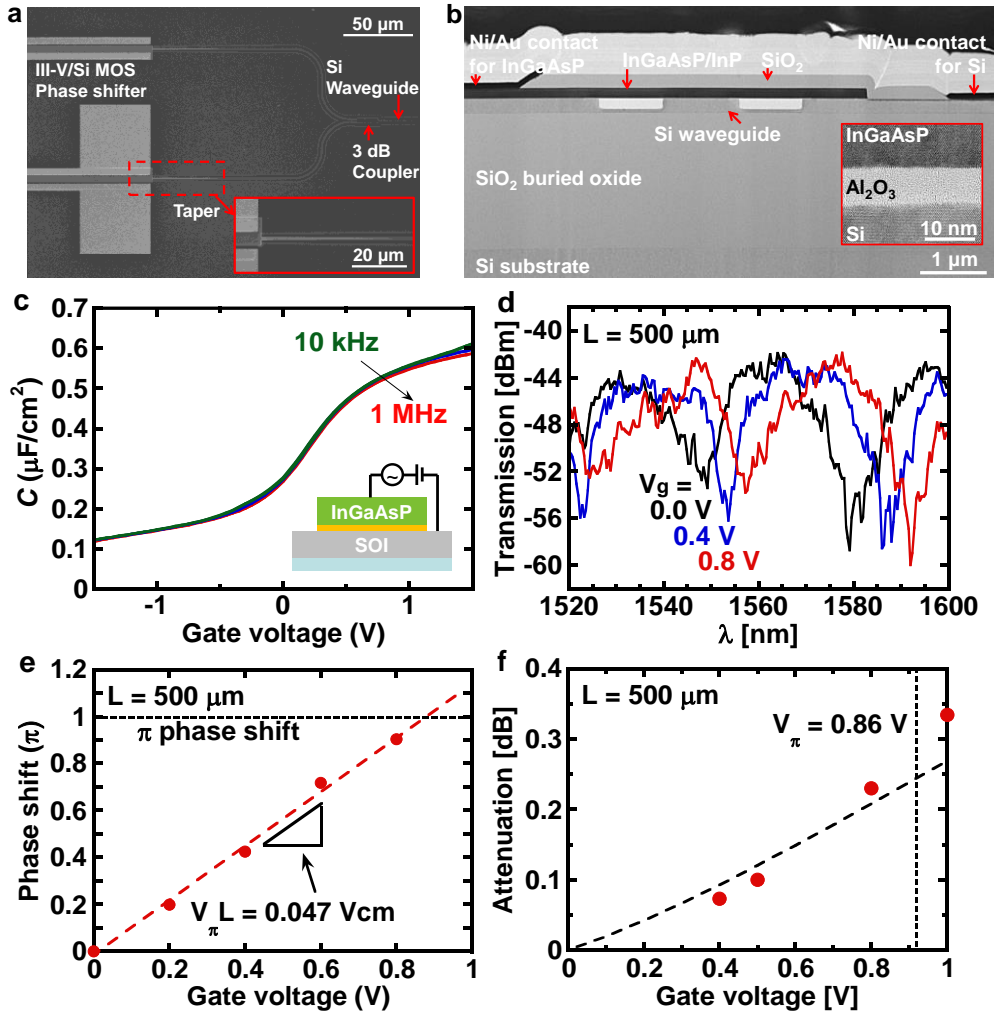

Fig. 3. **Fabrication and measurement of an InGaAsP/Si hybrid MOS optical modulator. a,** Scanning electron microscope image of the fabricated device. Inset: Enlarged view of a taper structure. **b,** Transmission electron microscope image of the fabricated device. Inset: Enlarged view of the wafer-bonded InGaAsP/Al₂O₃/Si MOS structure. **c,** Capacitance–voltage characteristic of the wafer-bonded InGaAsP/Al₂O₃/Si MOS capacitor. **d,** Transmission spectra of the asymmetric Mach–Zehnder interferometer with the InGaAsP/Si hybrid MOS phase shifter. **e,** Relationship between gate voltage and phase shift of an InGaAsP/Si hybrid MOS optical modulator. **f,** Attenuation increase of an InGaAsP/Si hybrid MOS phase shifter.

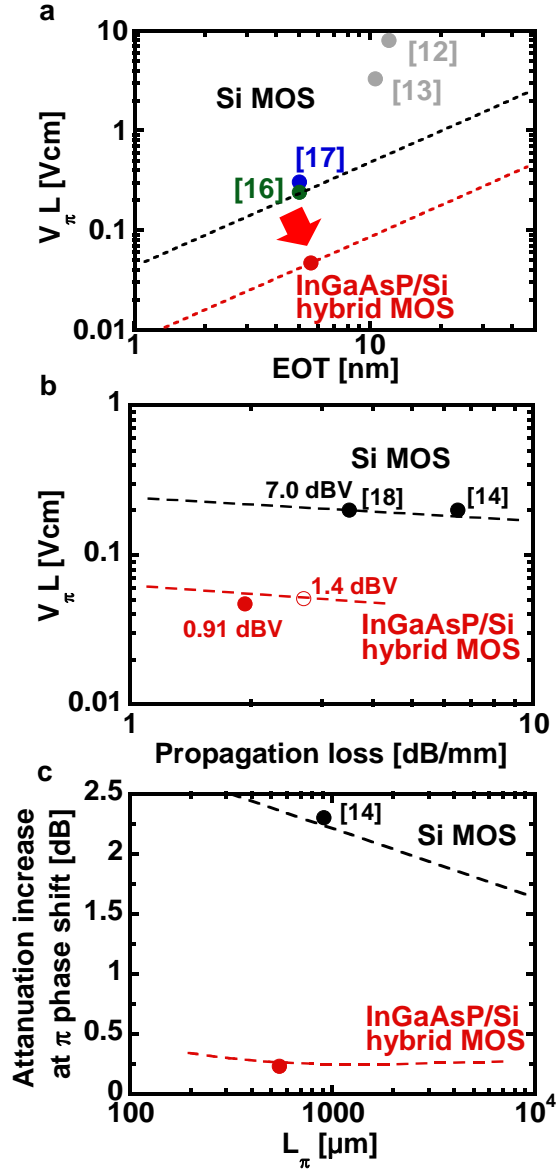

Fig. 4. **Benchmarks of the modulation efficiency and loss for an InGaAsP/Si hybrid MOS optical modulator. a,** Benchmark voltage–length product ($V_\pi L$) of MOS optical modulators as a function of EOT. **b,** Relationship between $V_\pi L$ and propagation loss of MOS optical modulators. **c,** Attenuation increase at $\pi$ phase shift of InGaAsP/Si hybrid and Si MOS phase shifters as a function of the phase shifter length ($L_\pi$). The dashed lines show simulated results and the closed circles show experimental results. The open circle shows an extrapolated value.

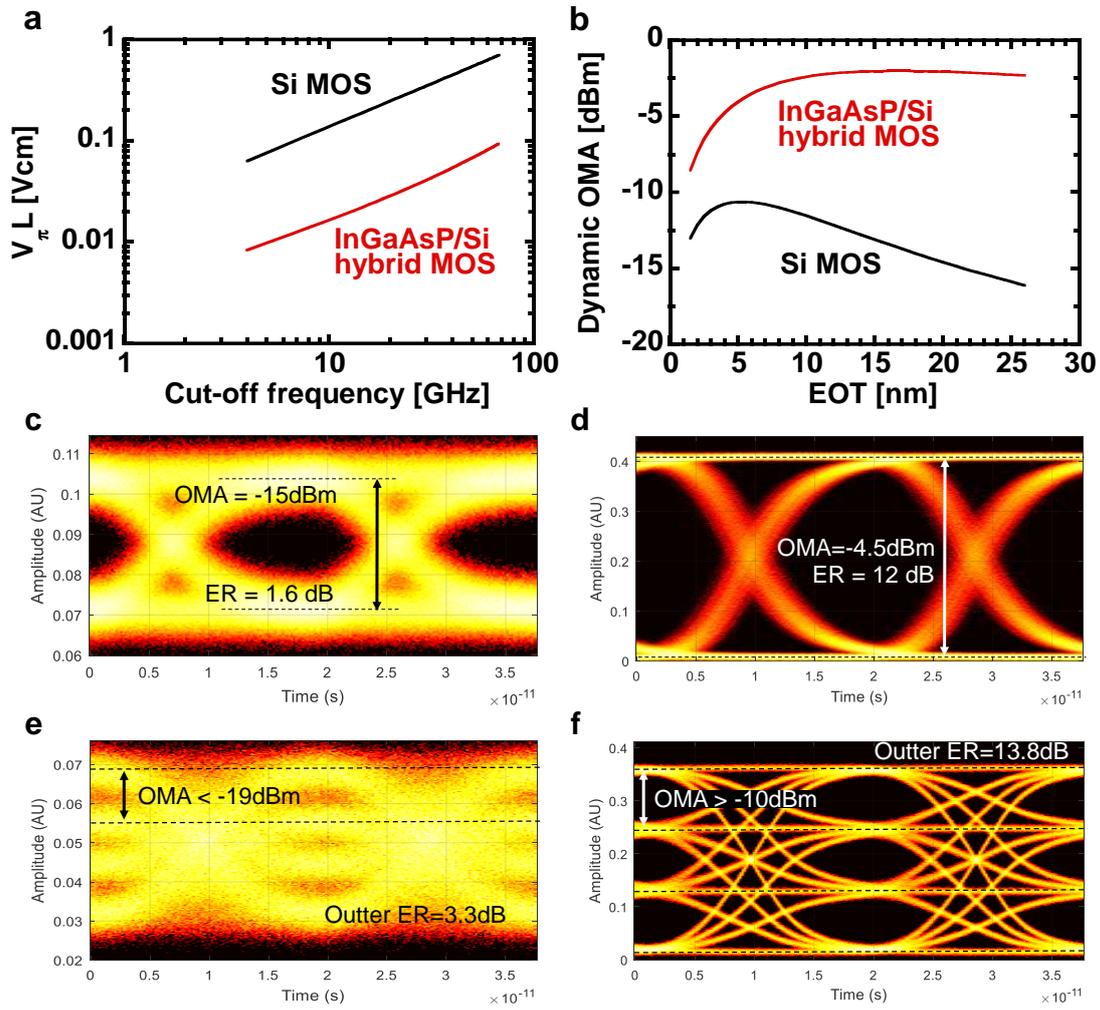

Fig. 5. **Benchmarks of the modulation bandwidth for an InGaAsP/Si hybrid MOS optical modulator. a,** Relationship between the modulation efficiency and cut-off frequency. **b,** Relationship between the dynamic OMA and the EOT. Eye patterns of 53-Gb/s NRZ for **(c)** a Si MOS optical modulator and **(d)** an InGaAsP/Si hybrid MOS optical modulator. Eye patterns of 53-Gb/s PAM-4 for **(e)** a Si MOS optical modulator and **(f)** an InGaAsP/Si hybrid MOS optical modulator.

**Supplementary information**

**High-modulation-efficiency InGaAsP/Si hybrid MOS optical modulator with**

**Mach–Zehnder interferometer**

Jae-Hoon Han, Frederic Boeuf, Shinichi Takagi, and Mitsuru Takenaka

The University of Tokyo, 7-3-1 Hongo, Bunkyo-ku, Tokyo 113-8656, Japan

E-mail: hanjh@mosfet.t.u-tokyo.ac.jp

## Contents

**I.   Taper structure for mode transition from Si waveguide to InGaAsP/Si hybrid MOS phase shifter**

The optical modes of a Si passive waveguide and an InGaAsP/Si hybrid MOS phase shifter are different because the total thickness of the phase shifter is thicker than the Si passive waveguide. For a smooth mode transition from the Si passive waveguide to the InGaAsP/Si hybrid MOS phase shifter, a suitable taper structure is required[1]. We designed III-V layers tapering on a Si waveguide. Figures S1a and S1b show the finite-difference time-domain (FDTD) simulation results of the mode propagation from the Si waveguide to the MOS phase shifter without and with an InGaAsP taper, respectively. When there was no taper, the multiple modes of the phase shifter were excited, resulting in increasing insertion loss. When the width of the 50-μm-long taper was gradually increased from 0.2 to 1 μm, we successfully obtained a smooth mode transition as shown in Fig. S1b.

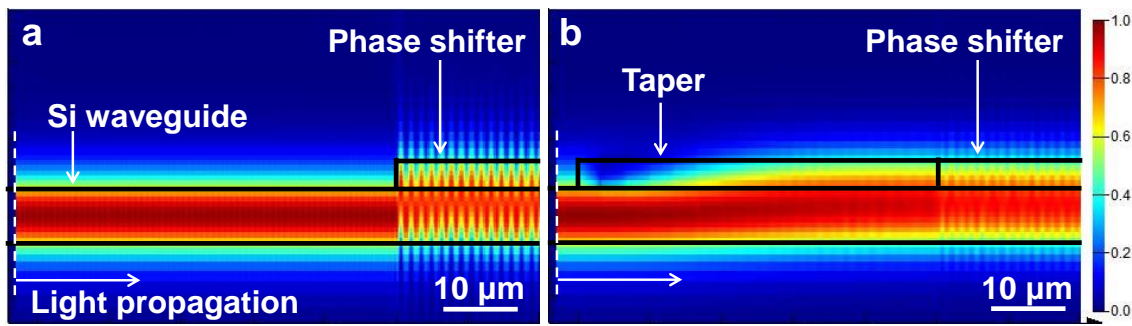

Fig. S1. **FDTD simulation of mode transition from a Si passive waveguide to an InGaAsP/Si hybrid MOS phase shifter.** Cross-sectional views of optical confinement calculated by FDTD simulation (**a**) without and (**b**) with a taper structure.

This taper structure also has an InGaAsP/Si hybrid MOS junction, which also contributes to phase modulation; thus, to calculate the phase modulation efficiency, we must take the effective length of the taper into account. To calculate this effective length, we considered the optical confinement in the region near the MOS interface at each position of the taper, which determined the interaction between the carriers and the propagating light. Figure S2 shows the normalized optical intensity confined in a 3-nm-thick region at the InGaAsP MOS interface as a function of the propagation length ($x$) calculated by FDTD simulation. The intensity was normalized by the intensity at $x$ of 50 μm. By integrating the normalized intensity multiplied by $x$, the effective taper length was found to be 24.4 μm. Therefore, we added this length to the length of the phase shifter to estimate the modulation efficiency more accurately.

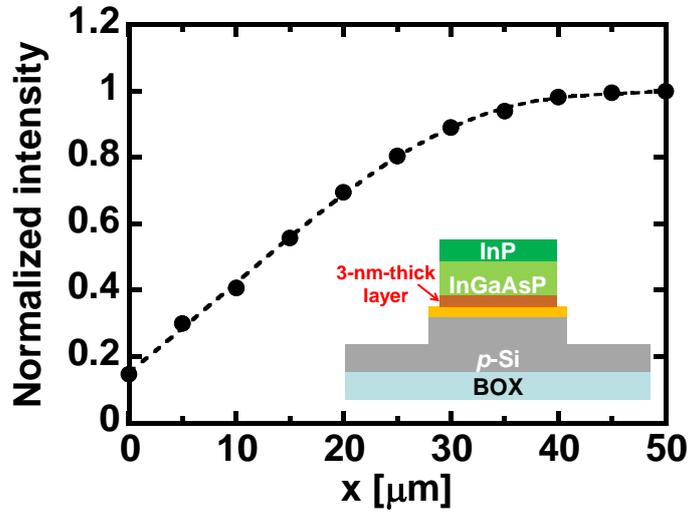

Fig. S2. **Normalized optical intensity of 3-nm-thick layer at InGaAsP MOS interface.** The optical intensity in the 3-nm-thick region at the InGaAsP MOS interface normalized by the intensity at $x$ of 50 μm as a function of the propagation length $x$ from the taper edge. From this result, the effective taper length was found to be 24.4 μm.

## II. Mode confinement of InGaAsP/Si hybrid MOS phase shifter

In the case of conventional Si hybrid optical modulators[2–4], the optical mode is not strongly confined because the height of the III-V layers is over 1 μm to avoid optical absorption caused by the top metal electrode; thus, the modulation efficiency is degraded owing to the weak optical confinement. We can partially mitigate this absorption by inserting an InGaAsP layer. Since the refractive index of InGaAsP is higher than that of InP, the optical confinement in the InP/InGaAsP layers (Fig. 2b) was greater than that in the single InP layer (Fig. 2a). To achieve further improvement in the optical confinement, we propose to use thin III-V layers as shown in Fig. S3c. By forming an electrical contact at the side of the III-V layers, we can completely eliminate the absorption of the metal even with such thin III-V layers.

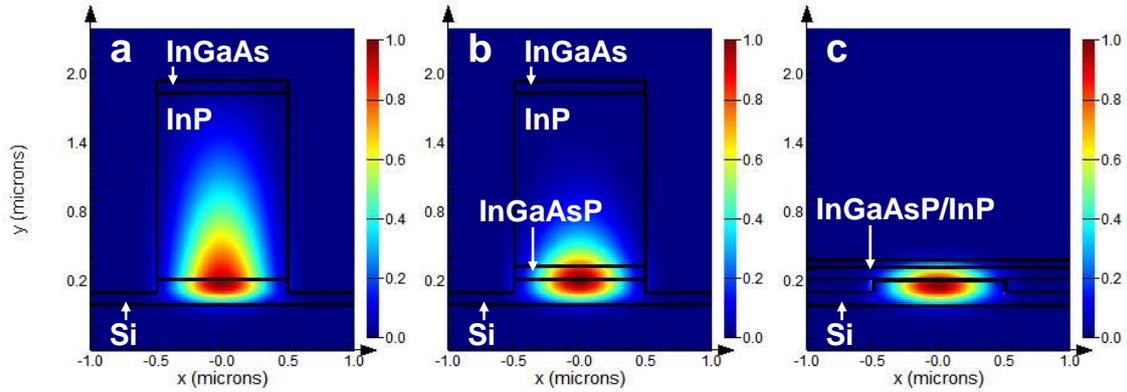

Fig. S3. **Optical modes of III-V/Si hybrid MOS phase shifter. a,** Optical mode of 100-nm-thick InGaAs/1.61-μm-thick InP/5-nm-thick $SiO_2$/220-nm-thick Si structure. **b,** Optical mode of 100-nm-thick InGaAs/1.5-μm-thick InP/110-nm-thick InGaAsP/5-nm-thick $SiO_2$/220-nm-thick Si structure. **c,** Optical mode of 50-nm-thick InP/110-nm-thick InGaAsP/5-nm-thick $SiO_2$/220-nm-thick Si structure. The width of the mesa structure was 1.0 μm.

## III. Resistivity reduction using crystalline *n*-type InGaAsP

Since *n*-InGaAsP has a larger electron mobility than Si[5], a reduction in resistivity is expected. To evaluate the resistivity of a Si-doped *n*-InGaAsP layer with a doping concentration of $10^{17}$ cm$^{-3}$ and the contact resistance between Ni-InGaAsP alloy[6,7] and an

*n*-InGaAsP layer, a transfer length measurement (TLM) was carried out for a wafer-bonded *n*-InGaAsP layer, as shown in Fig. S4a. From the TLM result in Fig. S4b, the resistivity and contact resistance were extracted to be $2.7 \times 10^{-3}$ $\Omega$cm and $1.2 \times 10^{-4}$ $\Omega$cm$^2$, respectively. The resistivity of the *n*-InGaAsP layer was 32 times lower than that of *n*-Si with the same doping concentration. This means that the modulation bandwidth, which was limited by the resistive-capacitance delay, can be improved by replacing the *n*-poly-Si with *n*-InGaAsP. Since the contact resistance between the Ni-InGaAsP alloy and the *n*-InGaAsP was sufficiently low, an ohmic contact to the *n*-InGaAsP was successfully formed even by a low-temperature process[6,7].

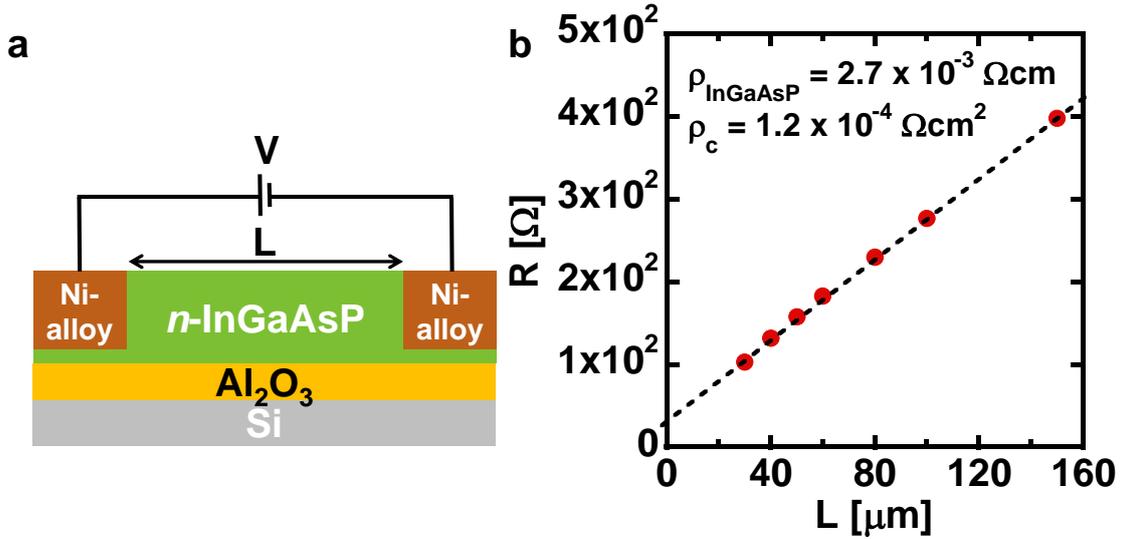

Fig. S4. **TLM measurement of InGaAsP. a,** Cross-sectional view of the TLM device. **b,** TLM measurement result.

## IV. Electron-induced changes in refractive index and absorption coefficient in InGaAsP

To analyze the characteristics of an InGaAsP/Si hybrid MOS optical modulator, the carrier-induced changes in the refractive index and absorption of $In_{1-x}Ga_xAs_yP_{1-y}$ were calculated. In contrast to Si, in which the plasma dispersion effect and free-carrier absorption are dominant[8], the band-filling effect, bandgap shrinkage, and inter-valence band absorption also contribute to the carrier-induced changes in the refractive index and absorption in InGaAsP[9.10].

To estimate the optical parameters of InGaAsP, its physical parameters taken from the literature were used. In the case of $In_{1-x}Ga_xAs_yP_{1-y}$ lattice-matched to InP, $x$ and $y$ have the relationship[11]

$x = 0.47y.$ (1)

The direct band gap of InGaAsP at the $\Gamma$ point is represented as[12]

$E_g = 1.35 - 0.72x + 0.12y^2$ [eV]. (2)

The band discontinuity between InP and InGaAsP[11] and the electron affinity of InP[13] are expressed by

$\Delta E_v = 502y - 152y^2$ [meV], (3)

$\Delta E_c = 268y - 3y^2$ [meV], (4)

$\chi_{InP} = 4.40$ [eV], (5)

where $\Delta E_v$ is the band offset of the valence band, $\Delta E_c$ is the band offset of the conduction band, and $\chi_{InP}$ is the electron affinity of InP. The electron affinity of InGaAsP can be calculated using $\chi_{InP}$ and $\Delta E_c$.

The electron and hole effective masses are[14]

$m_e = 0.08 - 0.039y$ [$1/m_0$], (6)

$m_{hh} = 0.56 - 0.22y + 0.11y^2$ [$1/m_0$], (7)

$m_{lh} = 0.12 - 0.092y + 0.024y^2$ [$1/m_0$], (8)

where $m_e$ is the electron effective mass at $\Gamma$ point, $m_{hh}$ is the heavy-hole effective mass, $m_{lh}$ is the light-hole effective mass, and $m_0$ is the electron rest mass. Figures S5a and 5b show the band gap of In$_{1-x}$Ga$_x$As$_y$P$_{1-y}$ lattice-matched to InP and its conductivity effective mass, respectively. The conductivity effective mass of electron $m_{ce}$ is considered to be the same as the effective mass of an electron at the Γ point. The conductivity effective mass of a hole $m_{ch}$ is represented as

$$m_{ch} = \frac{m_{hh}^{1.5} + m_{lh}^{1.5}}{m_{hh}^{0.5} + m_{lh}^{0.5}}. \text{ (9)}$$

The refractive index of InGaAsP is obtained using the model given by Adachi[15]. Figure S5c shows the relationship between the refractive index and the composition of As at a 1.55 μm wavelength. Some of the parameters needed for the calculation were obtained from the appendix of Weber's paper[10].

The electron and hole mobilities of InGaAsP were obtained by the model given by Sotoodeh *et al.*[5]. Figure S6 shows the estimated electron and hole mobility of InGaAsP for various donor and acceptor concentrations, respectively.

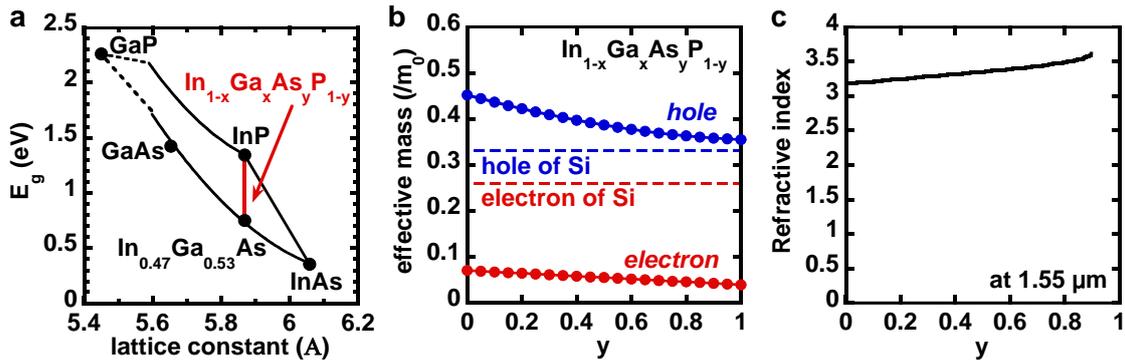

Fig. S5. **Physical parameters of In$_{1-x}$Ga$_x$As$_y$P$_{1-y}$ lattice-matched to InP. a,** Band gap of various III-V materials as a function of lattice constant[16]. **b,** Conductivity effective mass of InGaAsP[14]. **c,** Refractive index of InGaAsP at 1.55 μm[15].

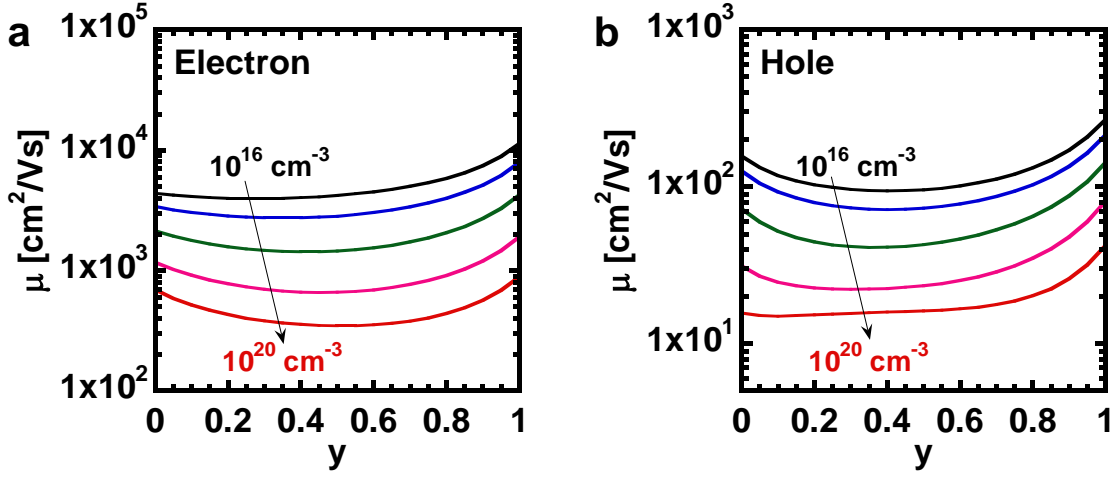

Fig. S6. **Carrier mobility of InGaAsP for various impurity concentrations[5]. a,** Electron mobility. **b,** Hole mobility.

According to the Drude model, the changes in the refractive index ($\mathit{\Delta}n$) due to the plasma dispersion effect and the changes in the absorption coefficient ($\mathit{\Delta}\alpha$) due to the free-carrier absorption are expressed by

$$\Delta n = -\frac{e^2 \lambda^2}{8\pi^2 c^2 \varepsilon_0 n}\left(\frac{\Delta N_e}{m_{ce}} + \frac{\Delta N_h}{m_{ch}}\right), (10)$$

$$\Delta \alpha = \frac{e^3 \lambda^2}{4\pi^2 c^3 \varepsilon_0 n}\left(\frac{\Delta N_e}{m_{ce}^2 \mu_e} + \frac{\Delta N_h}{m_{ch}^2 \mu_h}\right), (11)$$

where $e$ is the elementary charge, $\lambda$ is the wavelength, $c$ is the speed of light in vacuum, $\varepsilon_0$ is the permittivity in vacuum, $n$ is the unperturbed refractive index, $\mu_e$ is the electron mobility, and $\mu_h$ is the hole mobility. In the case of an $n$-type semiconductor, the effect of holes is negligible; thus, only the effect of electrons is considered in the equations.

The absorption change caused by the band-filling effect and bandgap shrinkage at photon energy $E$ is represented as[9,10]

$$\Delta\alpha(E) =$$

$$
\begin{cases}
0 & (E < E_{g,eff}) \\
\frac{C_{hh}}{E}\sqrt{E - E_{g,eff}}[f_v(E_{vh}) - f_c(E_{ch})] + \frac{C_{lh}}{E}\sqrt{E - E_{g,eff}}[f_v(E_{vl}) - f_c(E_{cl})] & (E_{g,eff} \le E \le E_g) \\
\frac{C_{hh}}{E}\sqrt{E - E_{g,eff}}[f_v(E_{vh}) - f_c(E_{ch})] + \frac{C_{lh}}{E}\sqrt{E - E_{g,eff}}[f_v(E_{vl}) - f_c(E_{cl})] - \alpha_0(E) & (E > E_g)
\end{cases}
$$

, (12)

$$
\alpha_0(E) = \frac{C_{hh}}{E}\sqrt{E - E_g} + \frac{C_{lh}}{E}\sqrt{E - E_g}, \text{ (13)}
$$

where $C_{hh}$ and $C_{lh}$ are proportional constants related to the density-of-states effective mass of heavy holes and light holes[10], and $f_v$ and $f_c$ are functions of the Fermi–Dirac distribution for holes and electrons, respectively. $E_{vh}$, $E_{vl}$, $E_{ch}$, and $E_{cl}$ are the energy levels of electron or heavy (or light) holes considered to energy and momentum conservation, where carriers can move between the two levels by the photon energy[9]. $E_{g,eff}$ is the effective band gap reduced by bandgap shrinkage, which lowers the energy of the conduction band edge when electrons exist at the conduction band[17]. The bandgap shrinkage is expressed by[17]

$$
\Delta E_g = \frac{An^{*-0.19} + Bn^{*1/3}}{1 + n_0/n^*}\frac{m_e^{1/2}}{\varepsilon_r}, \text{ (14)}
$$

where $A$ is $1.04 \times 10^3$, $B$ is $2.80 \times 10^{-7}$, $n_0$ is $2.40 \times 10^{19}$, $n^*$ is $\Delta N_e/m_e^{3/2}$, and $\varepsilon_r$ is the relative dielectric constant.

In the case of $p$-type InGaAsP, we considered the intervalence band absorption occurring between the valence band and the spin-off band[9,10].

Using these carrier-induced absorption changes, we calculated the carrier-induced refractive index change using the Kramers–Kroenig relation[9,10].

Figures S7a and S7b show the contributions of the plasma dispersion effect and the other effects including the band-filling effect and bandgap shrinkage to the electron-induced refractive index change in InGaAsP and InP, respectively. The wavelength ($\lambda_g$) corresponding to the bandgap energy of InGaAsP was assumed to be 1.37 μm.

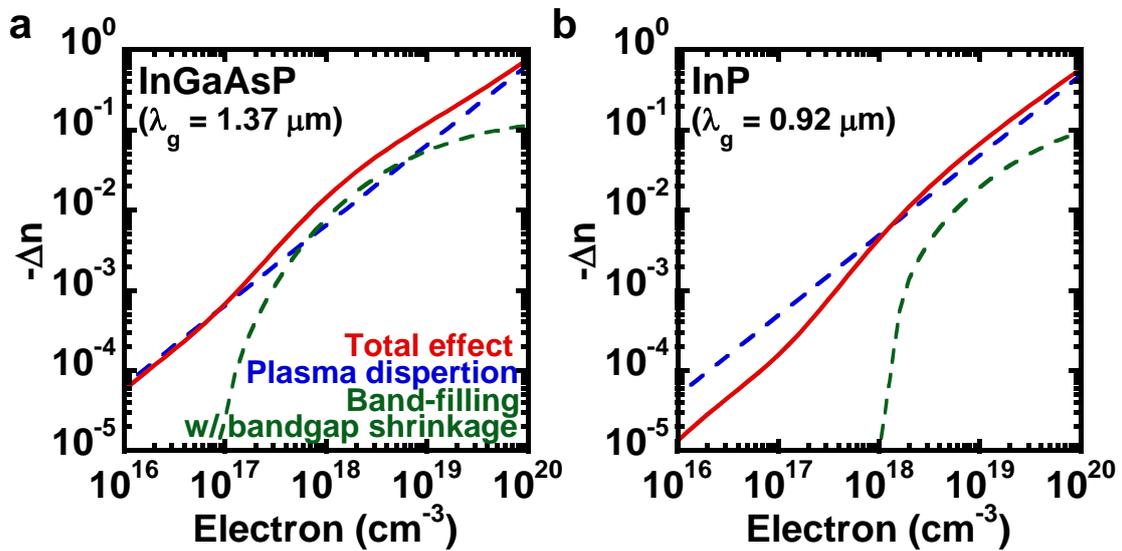

Fig. S7. **Details of electron-induced refractive index changes in InGaAsP and InP.** Electron-induced refractive index changes of (**a**) InGaAsP and (**b**) InP.

Figure S8 shows the hole-induced changes in the refractive index and absorption of InGaAsP, InP, and Si. InGaAsP was predicted to have an up to 3.6 times higher hole-induced refractive index change than Si, while the hole-induced absorption of InGaAsP was predicted to be 4.4 times larger than that of Si owing to the intervalence band absorption.

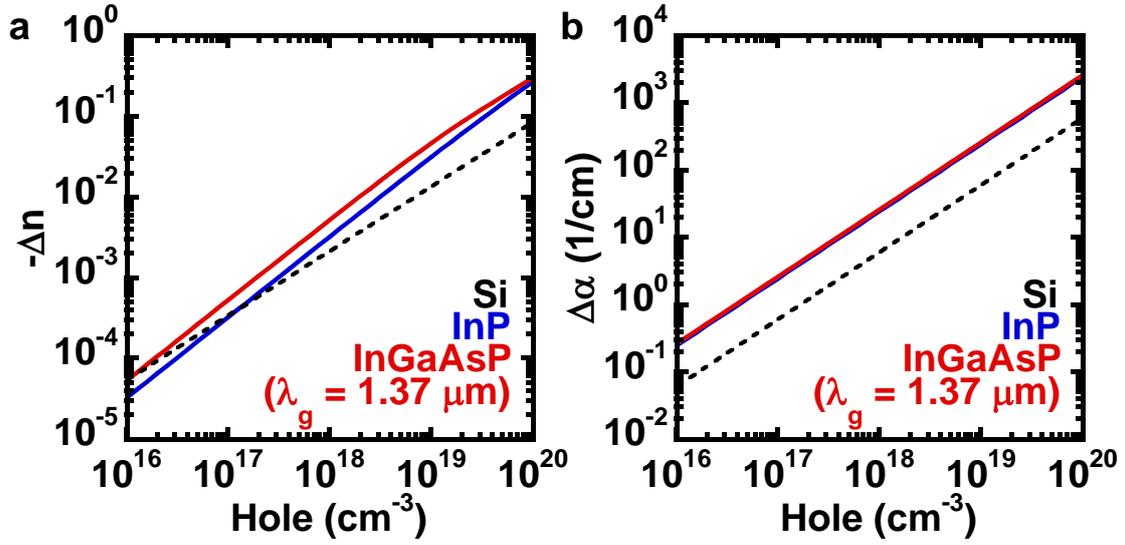

Fig. S8. **Hole-induced refractive index changes and absorption coefficient changes of Si, InP, and InGaAsP. a,** Hole-induced refractive index changes. **b,** Hole-induced absorption change.

To achieve high modulation efficiency with a small insertion loss for optical modulators, the ratio of the refractive index change ($\Delta n$) to the absorption coefficient change ($\Delta k$) is essential. Figures S9a and S9b show $\Delta n/\Delta k$ in InGaAsP and Si as a function of the electron or hole density, respectively. As shown in Fig. S9a, the modulation by electrons is more effective than that by holes in the case of InGaAsP. On the other hand, the modulation by holes is more effective than that by electrons in the case of Si as shown in Fig. S9b. Since the presented InGaAsP/Si hybrid MOS optical modulator can utilize the electron effect in InGaAsP and the hole effect in Si, we can simultaneously achieve high modulation efficiency and low insertion loss.

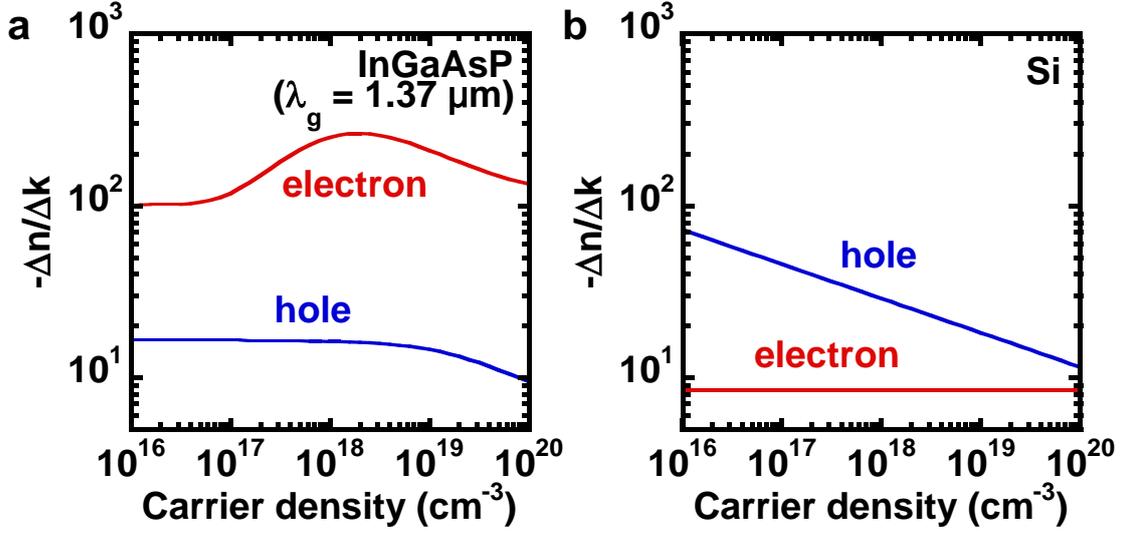

Fig. S9. **Ratio of -Δ*n* to Δ*k*.** The ratio is the figure of merit of the efficiency of the phase shift caused by carriers. **a,** -Δ$n$/Δ$k$ for InGaAsP. **b,** -Δ$n$/Δ$k$ for Si.

## V. Simulation of an InGaAsP/Si hybrid MOS optical modulator

The carrier distributions of an InGaAsP/Si hybrid MOS optical modulator were simulated by Synopsys technology computer-aided design (TCAD) Sentaurus. For accurate simulation, the energy levels of the conduction and valence band edges and the effective densities of states of the conduction band and valence band of InGaAsP and InP discussed in section IV were used. A schematic of the simulated structure is shown in Fig. S10a. The equivalent oxide thickness (EOT) of the $Al_2O_3$ gate oxide was assumed to be 5 nm, which is a typical value for state-of-the-art Si MOS optical modulators[8,9]. A gate voltage ($V_g$) was applied between the $n$-type InGaAsP and $p$-type Si layers. The simulated electron density distribution at $V_g$ of 1 V in Fig. S10b shows clear electron accumulation at the InGaAsP interface. Figures S10c shows the carrier distribution at the InGaAsP/Si hybrid MOS interface when the gate voltage is 0 and 1 V. This distribution shows the carrier density in the $y$ direction, as shown in Fig. S10b, where $x$ is 0 μm. It can be seen that electrons and holes accumulate at the InGaAsP and Si interfaces at 1 V, respectively. The band diagram at 1 V in Fig. S10d shows the large band bending at the InGaAsP interface exhibiting strong electron accumulation.

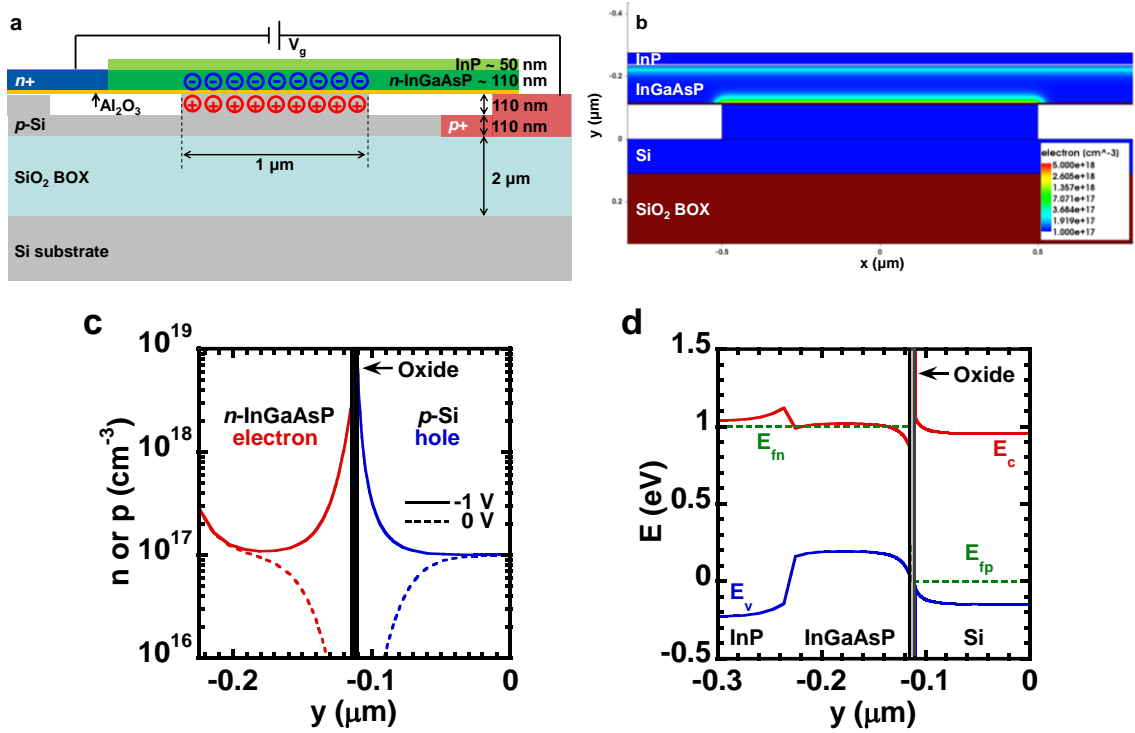

Fig. S10 **Simulation of an InGaAsP/Si hybrid MOS optical modulator. a,** Cross-sectional schematic of an InGaAsP/Si hybrid MOS optical modulator. **b,** Electron distribution at $V_g$ of 1 V. **c,** Carrier distribution of the simulated structure. **d,** Band diagram across the InGaAsP/Si hybrid MOS structure.

## VI. Fabrication procedure of InGaAsP/Si hybrid MOS optical modulator

The fabrication procedure is shown in Fig. S11. After boron implantation to form a $p$-Si region on a Si-on-insulator (SOI) wafer with a 220-nm-thick Si layer and 2-μm-thick buried oxide (BOX) layer, Si waveguides were formed by electron-beam lithography and inductively coupled plasma (ICP) etching. The doping concentration of $p$-Si was $3 \times 10^{17}$ cm$^{-3}$. Activation annealing for the boron was carried out at 1000 °C for 2 h. Then, an InP wafer containing InGaAsP/InP layers was bonded on the Si waveguide using an Al$_2$O$_3$ bonding interface. The doping concentration of in-situ Si-doped InGaAsP was $1 \times 10^{17}$ cm$^{-3}$. After removing the InP substrate and buffer layers, InGaAsP/InP mesas were formed by reactive ion etching (RIE) with CH$_4$/H$_2$ gases. After SiO$_2$ passivation and via formation, Ni was sputtered to form Ni-alloy contacts for the $n$-InGaAsP and $p$-Si layers, respectively. Ni-InGaAsP and Ni-Si alloys were formed by annealing at 400 °C for 1 min. Finally, Ni/Au pads were formed by EB evaporation and lift-off.

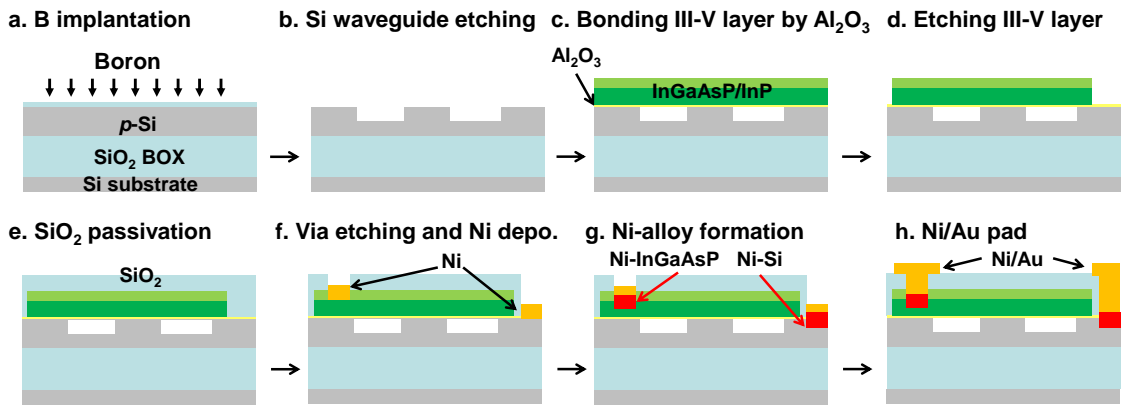

Fig. S11. **Process flow of InGaAsP/Si hybrid MOS optical modulator.**

## VII. Loss measurement of InGaAsP/Si hybrid MOS optical modulator

To measure the propagation loss of the InGaAsP/Si hybrid MOS phase shifter, we compared the optical output power between the Si passive waveguide and the InGaAsP/Si hybrid MOS phase shifter as shown in Figs. S12a and S12b, respectively. The phase shifter length without the length of the taper was 500 μm. Figure S12c shows the output power of each waveguide. The difference in the output power was caused by the InGaAsP/Si hybrid MOS structure; thus, the propagation loss of the 500-μm-long phase shifter was found to be 1.9 dB/mm.

To evaluate the attenuation increase at π phase shift, we measured the propagation loss of the InGaAsP/Si MOS phase shifter with various gate voltages. For this measurement, one of the arms of the asymmetric Mach-Zehnder interferometer was cut by focused ion beam etching. To obtain the attenuation increase, we subtracted the optical loss at 0 V from the optical loss at each gate bias.

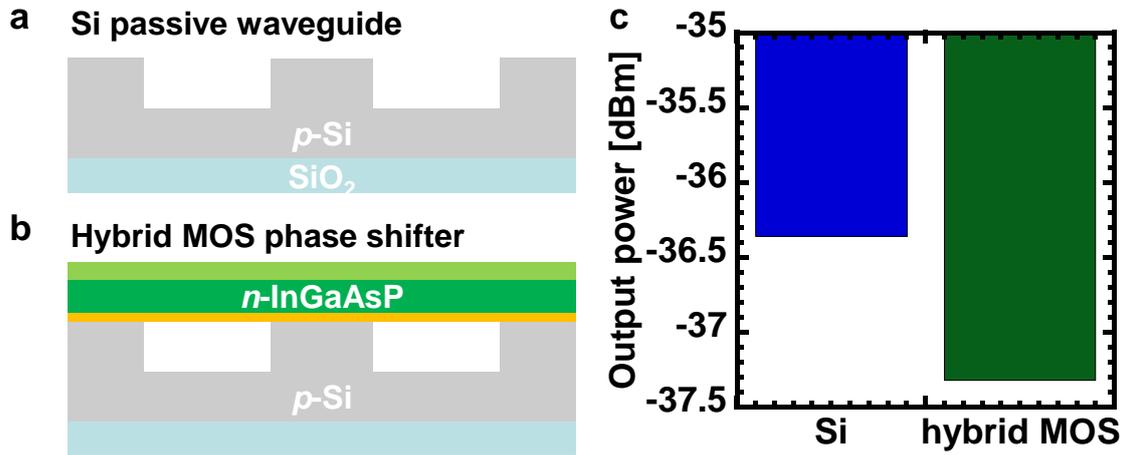

Fig. S12. **Measured propagation loss of InGaAsP/Si hybrid MOS optical modulator.** Cross-sectional views of **(a)** Si passive waveguide and **(b)** InGaAsP/Si hybrid MOS phase shifter. **c,** Optical loss of each waveguide.

## VIII. Analysis of dynamic modulation characteristics of an InGaAsP/Si hybrid MOS optical modulator

To discuss the modulation bandwidth of an InGaAsP/Si hybrid optical modulator, we considered the device structure as shown in Fig. S13a, which was similar to that of state-of-the-art Si MOS optical modulator reported in Refs. 18 to 20. The MOS structure with

*n*-type Si or InGaAsP on *p*-type Si was considered for numerical analysis. Figure S13b and c show the electrical field of a fundamental TE mode in a Si MOS optical modulator and an InGaAsP/Si hybrid MOS optical modulator, respectively.

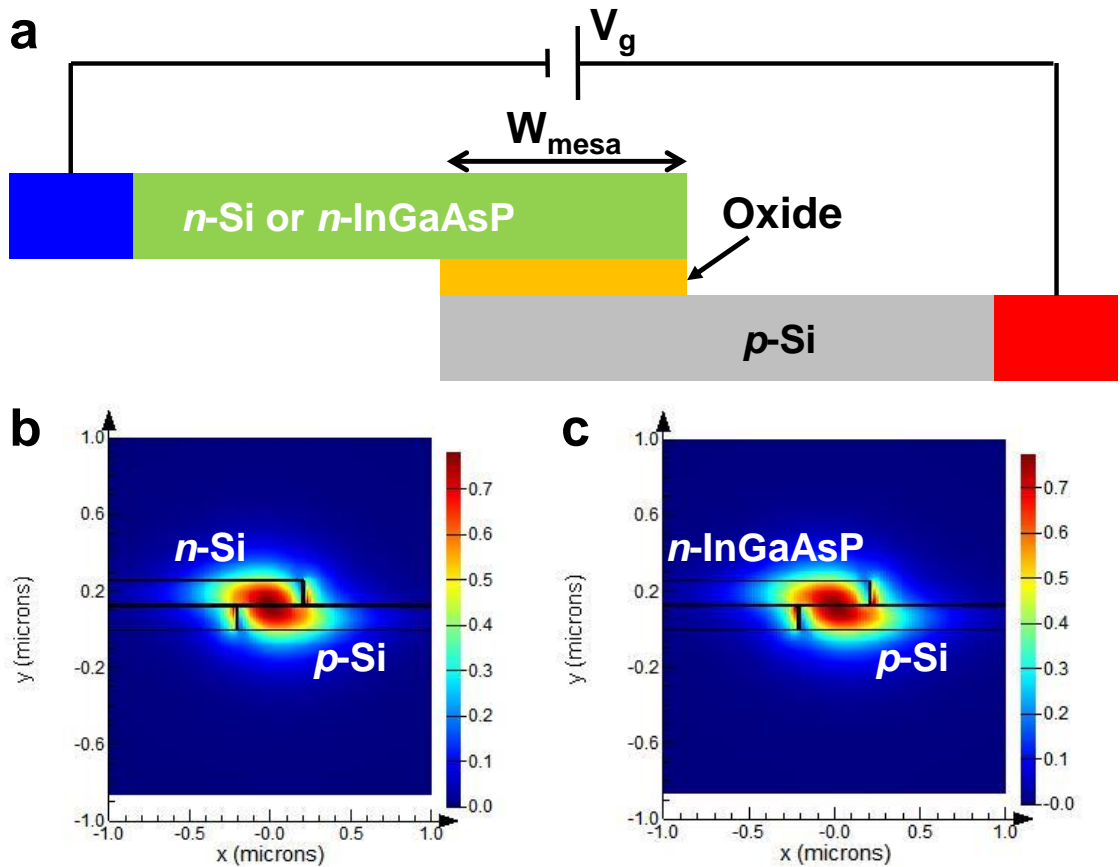

Fig. S13. **Structure for simulation of dynamic modulation characteristics. a,** Cross-sectional views of an InGaAsP/Si hybrid MOS and Si MOS optical modulator. Fundamental TE modes of (**b**) a Si MOS and (**c**) an InGaAsP/Si hybrid MOS optical modulator.

Overlap between accumulated carrier at the MOS junction and the optical mode was calculated by assuming a constant carrier density within an accumulation layer whose thickness was given by its Debye length. To confirm the accuracy of our model, we reproduced the result of state-of-the-art Si MOS optical modulator reported in Refs. 18 to 20. We estimated the following geometrical parameters of this Si MOS optical modulator: the waveguide width ($W_{mesa}$) was estimated to be 450 nm, the SOI and the poly-Si thicknesses to 120 nm, the doping concentration to be $1 \times 10^{17}$ cm$^{-3}$ in each Si layers, the oxide thickness to be 3.3 nm, and the flatband voltage ($V_{fb}$) to be -1.2 V. Comparison between simulated and experimental phase shift and loss dependences with gate voltage are shown in Fig. S14a and b, showing a very good agreement between our model and experimental data. Therefore, our numerical analysis is suitable to calculate the performance of an MOS optical modulator.

Then we reproduced eye diagram reported in Refs. 18 to 20 to evaluate dynamic modulation characteristic. In order to achieve 9 dB extinction ratio (ER) as experimentally obtained in Ref. 18, we estimated that the modulator total length was 360 µm, which was consistent with the layout data that can be guessed from Ref. 18. The driving voltage was set between $V_{fb}$ and $V_{fb}$ + 1 V, as detailed in Ref. 20 and the constant phase shift was set at a quadrature point. We also considered a 0 dBm available optical power prior to modulation and a propagation loss of 6.5 dB/mm at 0 V as indicated in Ref. 20. Only the segmented modulator design case was considered here. By assuming that the receiver stage had a typical high speed photodiode featuring a 3dB fiber-to-chip coupling loss, 1 A/W responsivity, and 10 nA dark current[21], we obtained the eye diagram modulated at 28 Gb/s non-return-to-zero (NRZ) signal as shown in Fig. 14c. Note that in the generated NRZ signal, 1.8 ps jitter and rise/fall time equal to 1/3 of the signal frequency were assumed. In addition, noise current from the photodiode was considered, while photodiode cut-off frequency was supposed to have no impact on the electrical response at the receiver side. The calculated ER of 9.5 dB and optical modulation amplitude (OMA) of 120 µW were in good agreement with data provided in Ref. 18 where ER was about 9.3 dB and OMA estimated to 105 µW.

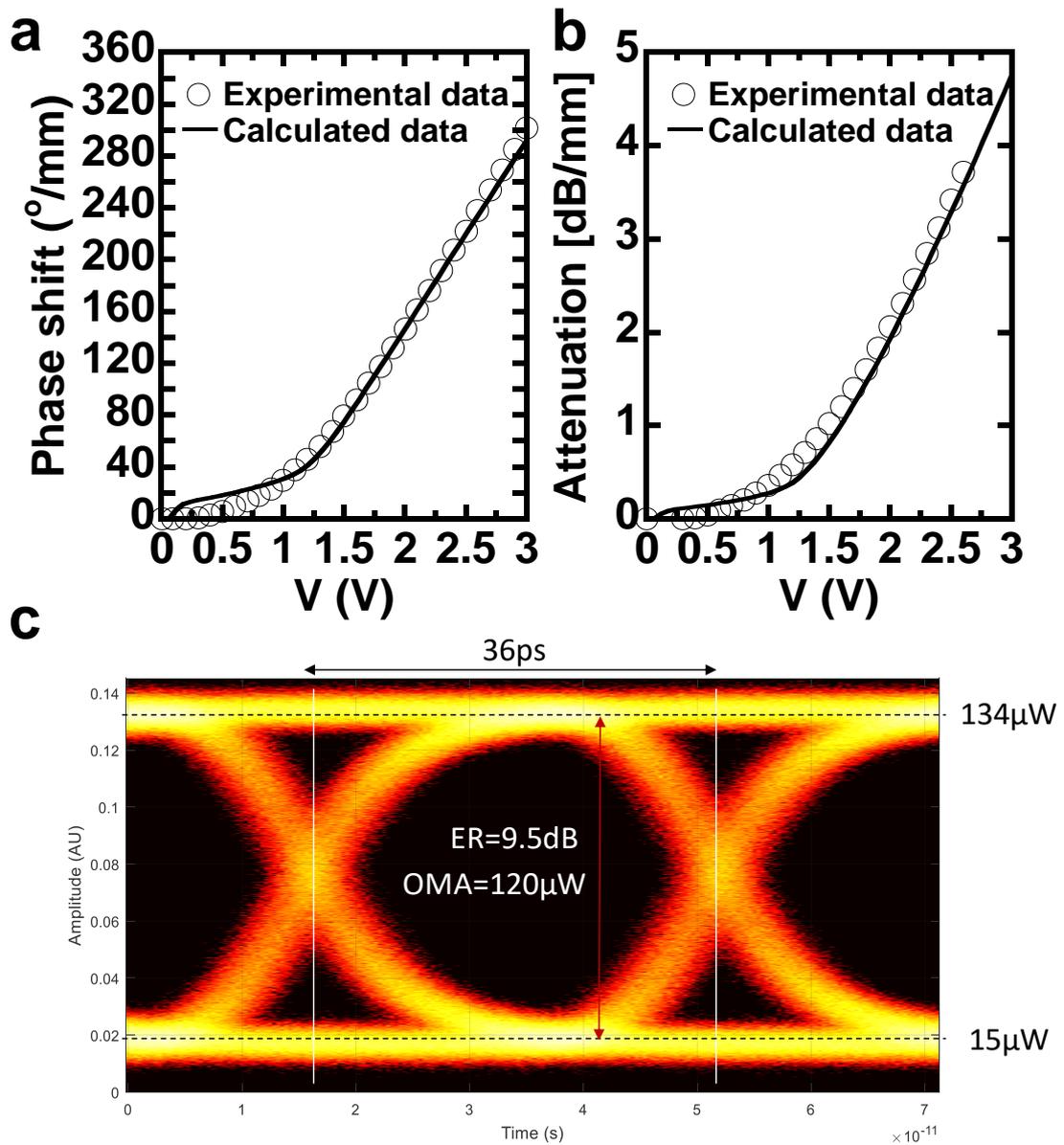

Fig. S14. **Calculated characteristics of a Si MOS optical modulator based on Cisco's structure.** Comparison between calculated data and reported data from Cisco in terms of **(a)** phase shift and **(b)** attenuation. **c,** Calculated eye diagram of a Si MOS optical

modulator at 28 Gb/s.

To reproduce an InGaAsP/Si hybrid MOS optical modulator by this method, the overlap between the optical mode and accumulated carrier at an InGaAsP/Si hybrid MOS structure was calculated by same method discussed above. Applying the developed model allowed reproducing well the behavior of an InGaAsP/Si hybrid MOS device as shown in Fig. S15, using the doping concentration of $3 \times 10^{17}$ cm$^{-3}$, $V_{fb}$ of -0.6 V and EOT of 5.7 nm extracted from experimental C-V curve (See Fig. 3c).

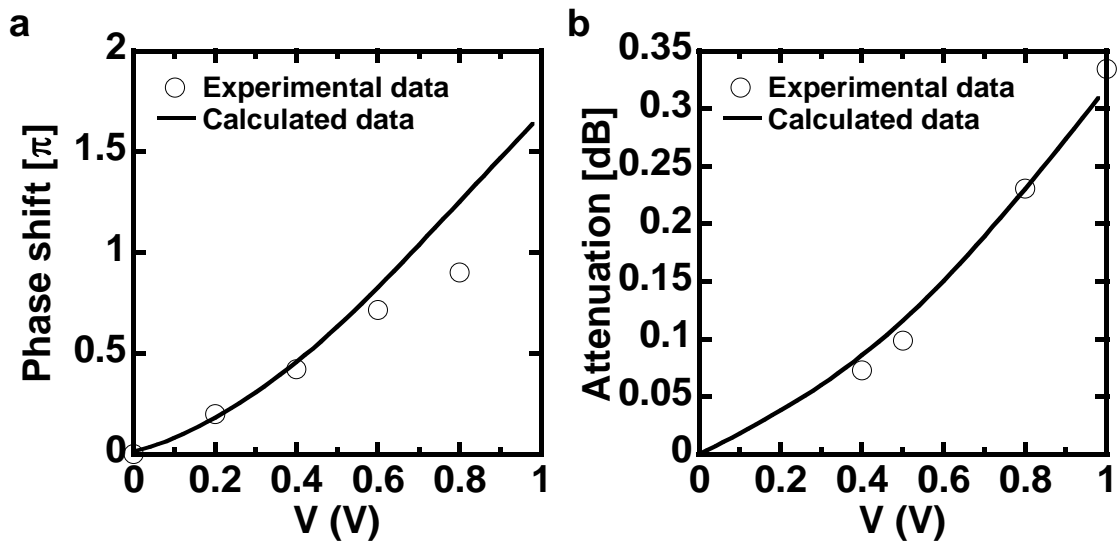

Fig. S15. **Calculated characteristics of an InGaAsP/Si hybrid MOS optical modulator.** Comparison between calculated data and experimental data in terms of **(a)** phase shift and **(b)** attenuation.

We have also calculated the eye diagram of an InGaAsP/Si hybrid MOS optical modulator where its $W_{mesa}$ is 420 nm. Figure S16 shows the eye diagram when the cut-off frequency was 25 GHz. Comparing with the eye diagram of Si MOS optical modulator as shown in Fig. S14c, it can achieve 3.5 times improvement in OMA and 12 times improvement in ER as compared with a Si MOS optical modulator, which is owing to the high modulation efficiency, small absorption, and reduced access resistance due to the low resistivity of *n*-InGaAsP layer.

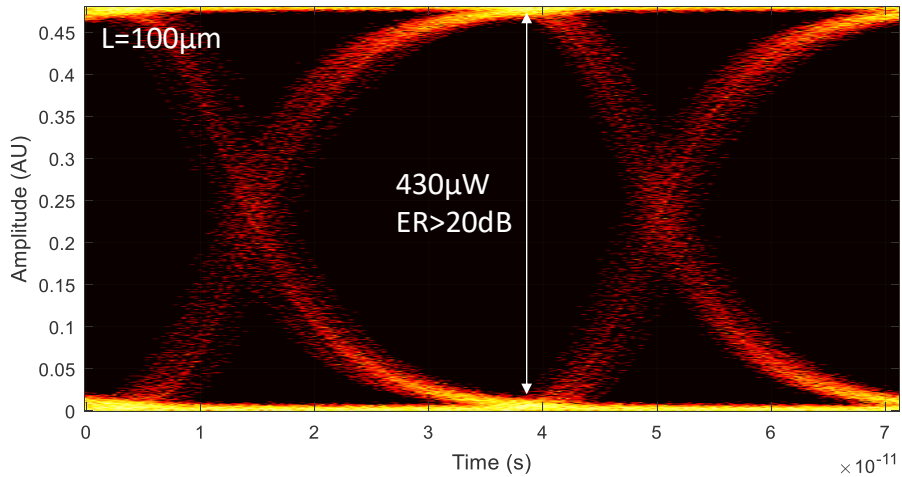

Fig. S16. **Calculated eye diagram of an InGaAsP/Si hybrid MOS optical modulator at 28 Gb/s.**

### References

1. Heck, M. J., Bauters, J. F., Davenport, M. L., Doylend, J. K., Jain, S., Kurczveil, G., Srinivasan, S., Tang, Y. & Bowers, J. E. Hybrid Silicon Photonic Integrated Circuit Technology. *IEEE J. Sel. Top. Quantum Electron.* **19**, 6100117 (2013).

2. Chen, H.-W., Kuo, Y.-H. & Bowers, J. E. A Hybrid Silicon–AlGaInAs Phase Modulator. *IEEE Photonics Technol. Lett.* **20**, 1920 (2008).

3. Kuo, Y.-H., Chen, H.-W. & Bowers, J. E. High speed hybrid silicon evanescent electroabsorption modulator. *Opt. Express* **16**, 9936 (2008).

4. Chen, H.-W., Peters, J. D. & Bowers, J. E. Forty Gb/s hybrid silicon Mach-Zehnder

modulator with low chirp. *Opt. Express* **19**, 1455 (2011).

5. Sotoodeh, M., Khalid, A. H. & Rezazadeh, A. A. Empirical low-field mobility model for III–V compounds applicable in device simulation codes. *J. Appl. Phys.* **87**, 2890 (2000).

6. Park, J.-K., Takenaka, M. & Takagi, S. Low resistivity lateral P–I–N junction formed by Ni–InGaAsP alloy for carrier injection InGaAsP photonic devices. *Jpn. J. Appl. Phys.* **55**, 04EH04 (2016).

7. Park, J.-K., Han, J.-H., Takenaka, M. & Takagi, S. InGaAsP variable optical attenuator with lateral PIN junction formed by Ni-InGaAsP and Zn diffusion on III-V on insulator wafer. *MRS advances* 1 (2016).

8. Soref, R. & Bennett, B. Electrooptical effects in silicon. *IEEE J. Quantum Electron.* **23**, 123–129 (1987).

9. Bennett, B. R., Soref, R. A. & Del Alamo, J. A. Carrier-induced change in refractive index of InP, GaAs, and InGaAsP. *IEEE J. Quantum Electron.* **26**, 113–122 (1990).

10. Weber, J.-P. Optimization of the carrier-induced effective index change in InGaAsP waveguides-application to tunable Bragg filters. *IEEE J. Quantum Electron.* **30**, 1801–1816 (1994).

11. Adachi, S. Physical Properties of III-V Semiconductor Compounds: InP, InAs, GaAs, GaP, InGaAs, and InGaAsP. (Wiley, New York, 1992).

12. Nahory, R. E., Pollack, M. A., Johnston Jr., W. D. & Barns, R. L., Band gap versus composition and demonstration of Vegard's law for $In_{1-x}Ga_xAs_yP_{1-y}$ lattice matched to InP. *Appl. Phys. Lett.* **33**, 659 (1978).

13. Fischer, T. E. Photoelectric Emission and Work Function of InP. *Phys. Rev.* **142**, 519 (1966).

14. Pearsall, T. P. GaInAsP Alloy Semiconductors (Wiley, New York, 1982).

15. Adachi, S. Material parameters of $In_{1-x}Ga_xAs_yP_{1-y}$ and related binaries. *J. Appl. Phys.* **53**, 8775 (1982).

16. Foyt, A. G. The electro-optic applications of InP. *J. Cryst. Growth* **54**, 1 (1981).

17. Botteldooren, D. & Baets, R. Influence of bandgap shrinkage on the carrier-induced refractive index change in InGaAsP. *Appl. Phys. Lett.* **54**, 1989 (1989).

18. Webster, M. *et al.* An efficient MOS-capacitor based silicon modulator and CMOS drivers for optical transmitters. *In Int. Conf. Group IV Photonics* Paper WB1 (IEEE, 2014).

19. Webster, M., *et al.* Silicon Photonic Modulator Based on a MOS-Capacitor and a CMOS Driver. *In Compound Semiconductor Integrated Circuit Symp.* Paper E.3 (IEEE, 2014).

20. Webster, M., *et al.* Low Power MOS Capacitor Based Silicon Photonic Modulators and CMOS Drivers. *In Optical Fiber Conf.* Paper W4H.3 (OSA, 2015).

21. Boeuf, F., Cremer, S., Temporiti, E., Fere, M., Shaw, M. Baudot, C., Vulliet, N., Pinguet, T., Mekis, A., Gianlorenzo, M., Petiton, H., Maitre, P. l., Traldi, M. & Maggi, L. Silicon Photonics R&D and Manufacturing on 300-mm Wafer Platform. *J. lightwave technol.* **34**, 286 (2016).